# A pipeline for matching bibliographic references with incomplete metadata: experiments with Crossref and OpenCitations

Matteo Guenci[1] [0009-0006-3139-1667, matteo.guenci2@unibo.it],
Ivan Heibi[1,2] [0000-0001-5366-5194, ivan.heibi2@unibo.it],
Chiara Parravicini[1] [0009-0006-7501-6789, chiara.parravicini@studio.unibo.it],
Silvio Peroni[1,2] [0000-0003-0530-4305, silvio.peroni@unibo.it],
Marta Soricetti[1] [0009-0008-1466-7742, marta.soricetti2@unibo.it]

[1]Research Centre for Open Scholarly Metadata, Department of Classical Philology and Italian Studies, University of Bologna, Bologna, Italy
[2]Digital Humanities Advanced Research Centre, Department of Classical Philology and Italian Studies, University of Bologna, Bologna, Italy

## Abstract

While Crossref makes available more than 1.8 billion bibliographic references from publications for which it provides a DOI, more than 698 million of these references do not specify a DOI, making the creation of a formal citation link from the citing entity and the cited entity problematic. In this article, we propose an analysis of Crossref bibliographic references to show how we can use the unstructured text defining such references and the available (and partial) metadata specified in them to (a) map them to existing entities included in OpenCitations Meta and, then, (b) to enable the potential inclusion of additional and valid citations link among these entities. We have defined a precise methodology to address the analysis and run it against a manually defined Gold Standard and a subset of Crossref. While the heuristic-based tool developed has demonstrated strong matching precision and effective metadata integration, its recall limitations highlight the necessity of further enhancements to address metadata inconsistencies and leverage additional sources of citation data.

## 1. Introduction

A few years ago, Shotton (2013) stressed that was uncomprehensible how, in a world where the open access to scholarly publication was becoming the norm, the citations created by the scholarly literature – a core aspect in scholarly communication at large which enable the attribution of credit and integrate scientists' independent research endeavours, and which may serve different purposes such as studying the dynamics of the scholarly domain (Fortunato et al., 2018) – were not recognised as a part of the commons and, thus, freely and legally available for sharing by anyone. He pushed for the creation of open repositories that host and provide these citations for any purpose, and he took the first steps in that direction by releasing the first version of the OpenCitations Corpus in 2010.

This was one of the glimmers that enabled the scholarly community to start a discussion and a process for changing the *status quo* of the availability of citation data at large, which, in 2013, was still primarily governed by proprietary services that offered such data only for consistent fees. To provide an order of magnitude about the costs, the Conference of Italian University Rectors (CRUI) has negotiated with Elsevier, in 2018, a contract of 8 million euros to enable Italian universities to access (via REST API) SCOPUS data for five years (2019-2023) and with Clarivate Analytics, in 2020, another contract of 10 million euros to access (via periodic ad-hoc dumps) Web of Science data for five years (2021-2025) – data available at https://www.crui.it/documenti/54/New-category/1317/Trasparenza-contratti-2023_gennaio-2024.xlsx.

Indeed, a few years later, in 2017, the Initiative for Open Citations (I4OC) was launched with the goal of pushing academic publishers – using Crossref (Hendricks et al., 2020) as DOI registration agency for their articles and, thus, depositing there their metadata – to release also the reference lists of their articles to be freely accessible via the Crossref API. While in 2017 that was a choice in charge of the publishers, since 2022 all the metadata and references deposited in Crossref are openly available to anyone without further, as explained in Crossref Blog (https://www.crossref.org/blog/amendments-to-membership-terms-to-open-reference-distribution-and-include-uk-jurisdiction/).

Since 2017, these citation data available in Crossref have been used by several research works, e.g. (Borrego et al., 2023), and infrastructures, e.g. the Ukrainian Open Citation Index (Cheberkus & Nazarovets, 2019), to analyse the scholarly domain and provide services to the community. OpenCitations (https://opencitations.net) (Peroni & Shotton, 2020), one of the organisations that collaborated to form I4OC, has been using bibliographic metadata and citation data from Crossref since 2015, when the new instance of the OpenCitations Corpus was released (Peroni et al., 2017). OpenCitations is a community-guided, non-profit Open Science infrastructure dedicated to publishing open bibliographic metadata and citation data. It hosts data from two different collections called OpenCitations Index (Heibi et al., 2024), which contains (as of 21 November 2025) more than 2.2 billion citation links among bibliographic resources, and OpenCitations Meta (Massari et al., 2024), which contains basic bibliographic metadata of all the citing and cited entities involved in the citations in the OpenCitations Index, extracted from five different sources – i.e. Crossref, DataCite, the National Institute of Health Open Citation Collection, OpenAIRE, and the Japan Link Centre.

While OpenCitations contains a huge amount of citation links, it does not include a citation for each bibliographic reference exposed in Crossref. Indeed, each bibliographic reference for a given article available in Crossref can be structured according to how the source provider (e.g., the publisher) decided to create the descriptive elements that define the reference. In particular, a bibliographic reference can be provided either as pure unstructured text or as a set of complete or incomplete metadata elements, each defined with appropriate fields. In case the field "doi" is not explicitly provided by the publisher for the bibliographic references of its articles, it may be provided by Crossref after the execution of automatic scripts that try to infer the correct DOI from a bibliographic reference.

The OpenCitations ingestion workflow for Crossref data considers only citations from bibliographic references that specify a DOI as a structured field in the JSON returned by the Crossref API. This means that a huge number of references (i.e., 698,198,538 out of 1,812,811,997 in the April 2025 Crossref dump) are simply discarded during ingestion.

Considering this scenario, in this article, we propose an analysis of Crossref bibliographic references to show how we can use the unstructured text defining such references and the available (and partial) metadata specified in them to (a) map them to existing entities included in OpenCitations Meta and, then, (b) to enable the potential inclusion of additional and valid citations link among these entities. In particular, we want to answer the following research questions (RQ1-RQ2):

1. How many Crossref references can be matched with the bibliographic resources included in OpenCitations – and, consequently, how many citations can we add to OpenCitations when the DOI of the cited entity is not specified in the original Crossref metadata?
2. What is the performance of the mapping, in particular considering the precision of the bibliographic resources recognised from the references and the potential loss?

To answer these questions, we have created a methodology implemented in a pipeline that combines state-of-the-art algorithms and tools and can potentially be used with other bibliographic and citation indexes different from OpenCitations'. However, in this article, we focus specifically on OpenCitations collections, using, as an experimental setting, a population of bibliographic reference data from Crossref to demonstrate the efficacy of the approach and the potential benefits it offers to OpenCitations and the community.

The rest of the paper is organised as follows. In Section 2, we introduce some of the most important related work on the topic, which we used as a starting point and main inspiration for developing the methodology introduced in Section 3. Section 4 introduces the analysis results, which will be discussed in depth in Section 5. Finally, Section 6 concludes the article by outlining future work.

## 2. Related works

While the increasing availability of open citation data has created unprecedented opportunities for bibliometric analysis, significant obstacles remain regarding data completeness, consistency, and interoperability. Oftentimes, methods for linking and matching bibliographic records rely heavily on unique identifiers, such as DOIs (Visser et al. 2021). However, a relevant portion of scholarly publications lacks a persistent identifier (Heibi et al. 2019; Martín-Martín et al. 2021). Besides the issue of missing identifiers, further challenges arise from inconsistencies in publication dates, different alphabets for titles and author names, variations in title formatting, and, more generally, the heterogeneity and incompleteness of metadata. For these reasons, the need to develop alternative matching strategies that can effectively link records even with incomplete or inaccurate metadata becomes more urgent. In this context, heuristic and machine-learning-based approaches have emerged as promising means for enabling more reliable entity resolution (Sefid, 2022).

Visser et al. (2021) provide a large-scale comparison of five major bibliographic databases, Scopus, Web of Science, Dimensions, Crossref, and Microsoft Academic, highlighting the issues of metadata completeness, interoperability, and coverage discrepancies. Their work is particularly relevant to the present study as it provides the heuristic cascading approach, designed as a cascading sequence of increasingly permissive criteria, which represents the starting point for developing our methodology for metadata matching. The authors developed a six-step hierarchical approach that prioritises precision in the initial stages while progressively increasing recall through more relaxed conditions:

1. **Strict DOI matching:** The first and most precise step involves direct DOI-based matching for documents with identical publication years.
2. **Volume- and page-based matching:** If no DOI match is found, records are compared by publication year, volume number, and page or article number.
3. **Author- and page-based matching:** In the absence of volume information, matching is attempted using publication year, the last name of the first author, and the page or article number.
4. **Author- and volume-based matching:** This step retains the first author's last name and volume number while relaxing page number constraints.
5. **Source title matching:** If the previous criteria fail, documents are matched using ISSN or ISBN, along with page numbers.
6. **Title similarity matching:** The final and least restrictive step employs a title-based similarity measure that identifies potential matches based on the longest shared words.

This cascading approach balances precision and recall by progressively introducing less restrictive conditions, thereby maximising the probability of capturing true positive matches while minimising false positives. To refine their matching procedure, Visser et al. (2021) implemented a scoring mechanism where each attribute match contributes to a cumulative similarity score. Matches that exceed a predefined threshold are considered valid, whereas those below it are discarded. The study reports that over 80% of matches were successfully identified in the first step (DOI matching), while subsequent steps accounted for progressively fewer matches. The final step, based on title similarity, was the least precise but proved crucial for capturing documents with incomplete metadata.

A cascading matching strategy was also adopted by Olensky et al. (2015) in their study, which provides a detailed evaluation of citation matching algorithms used by bibliometric research groups, focusing on the methods developed by the Centre for Science and Technology Studies (CWTS) and the Institute for Research Information and Quality Assurance (iFQ). Their work is particularly relevant to this research, as it investigates the impact of metadata inaccuracies on citation linkage and proposes strategies to mitigate these issues through heuristic and rule-based approaches. Olensky et al. (2015) identify two principal challenges in citation matching. First, metadata errors, such as variations in author names, journal abbreviations, and volume or page numbers, can prevent automated linking based on strict matching criteria. Second, citation indexes differ in how they extract and structure bibliographic metadata, leading to variations that complicate cross-database integration. To address these issues, CWTS and iFQ developed iterative, rule-based matching algorithms that employ a cascading sequence of increasingly permissive criteria.

The CWTS algorithm follows a hierarchical approach, beginning with strict rules that require exact matches for key bibliographic fields before progressively relaxing them. The algorithm starts with exact matching on the first author's last name, publication year, journal title, volume number, and starting page number. If no match is found, a second iteration allows for minor variations, such as a one-year difference in the publication year or alternative abbreviations of the journal title. Subsequent iterations incorporate fuzzy matching techniques, including Soundex encoding for author names and Levenshtein distance for journal titles, enabling the algorithm to tolerate minor spelling errors and formatting inconsistencies. The algorithm also prioritises matches based on citation frequency, selecting the most highly cited candidate when multiple matches are possible.

Similarly, the iFQ algorithm employs a structured set of rules that progressively relax matching criteria. Unlike the CWTS approach, which initially focuses on author and journal metadata, the iFQ method places greater emphasis on numerical fields such as volume and page numbers. The algorithm first attempts an exact match based on these attributes, then allows for small deviations, such as page number shifts or missing volume data. If necessary, additional iterations use Damerau-Levenshtein distance to compare journal titles and author names. Notably, the iFQ algorithm permits ambiguous matches, retaining multiple possible links when an exact resolution is not possible. This approach enhances recall but necessitates additional verification to resolve potential false positives.

A key finding of Olensky et al. (2015) is that iterative heuristic-based approaches significantly improve recall without compromising precision. The study reports that the CWTS and iFQ algorithms outperform the citation matching system used by Web of Science, which relies on a more rigid matching framework. While WoS achieves high precision, it fails to compensate for metadata inconsistencies, resulting in a substantial number of missed matches. In contrast, the adaptive heuristics employed by CWTS and iFQ successfully recover citations that would otherwise remain unlinked, demonstrating the effectiveness of flexible, metadata-tolerant matching strategies.

Hendricks et al. (2020) examine metadata quality within Crossref, highlighting its role as a central repository for scholarly identifiers. Although Crossref collects, preserves, and makes metadata available for public use, the accuracy and completeness of the data depend on member organisations that submit records. Missing fields, such as author affiliations, page numbers, and ISSNs, can interfere with accurate citation linking, as Crossref does not independently correct or edit submitted metadata. Corrections must be made by the original data providers, meaning inconsistencies persist unless actively updated by the contributing publishers.

To improve metadata completeness, Crossref employs enrichment techniques that integrate additional information into deposited records. One of the most significant improvements involves linking missing bibliographic references between documents through fuzzy comparisons with existing metadata in the system. This approach enhances discoverability and citation linking by inferring connections that may be incomplete or inconsistently formatted. Additionally, missing funder identifiers are inferred from the Open Funder Registry, ensuring that funding metadata is properly attributed even when not explicitly provided by the original data submitters.

Despite these efforts, metadata inconsistencies in Crossref remain an issue, as enrichment processes do not replace the need for accurate initial data submission. The findings of Olensky et al. (2015) and Hendricks et al. (2020) underscore the need for hybrid approaches that combine structured metadata integration with flexible matching techniques. The present study builds on these insights by refining bibliographic integration methodologies to enhance metadata quality and improve the accuracy of citation matching.

# 3. Methodology

Considering the state of the art and the problems posed by bibliographic metadata inconsistencies, OpenCitations constitutes an ideal case study. OpenCitations Meta aggregates data from multiple databases. Among these, Crossref is one of the main sources for OpenCitations, even if sometimes Crossref data lack complete metadata regarding references. This premise enables this study to examine the advantages and limitations of a heuristic approach that uses multiple metadata fields to match bibliographic references.

The methodology is based on four macro phases:

i.   data collection and preparation;

ii.  bibliographic metadata matching;

iii. enrichment fallback step for unstructured references;

iv. results exportation and evaluation.

These phases are described in the workflow diagram in Figure 1: each phase has a set of components grouped by a specific colour. The rest of this section will go into these phases in more detail. The last phase will be discussed in Section 4, when the results are presented. The methodological approach alongside the implementation process is introduced in the following subsections.

The tool running the methodology is available via Zenodo (Guenci, et al. 2025) to enable the reproducibility of the study. More details about these resources can be found in Section "Data Availability Statement".

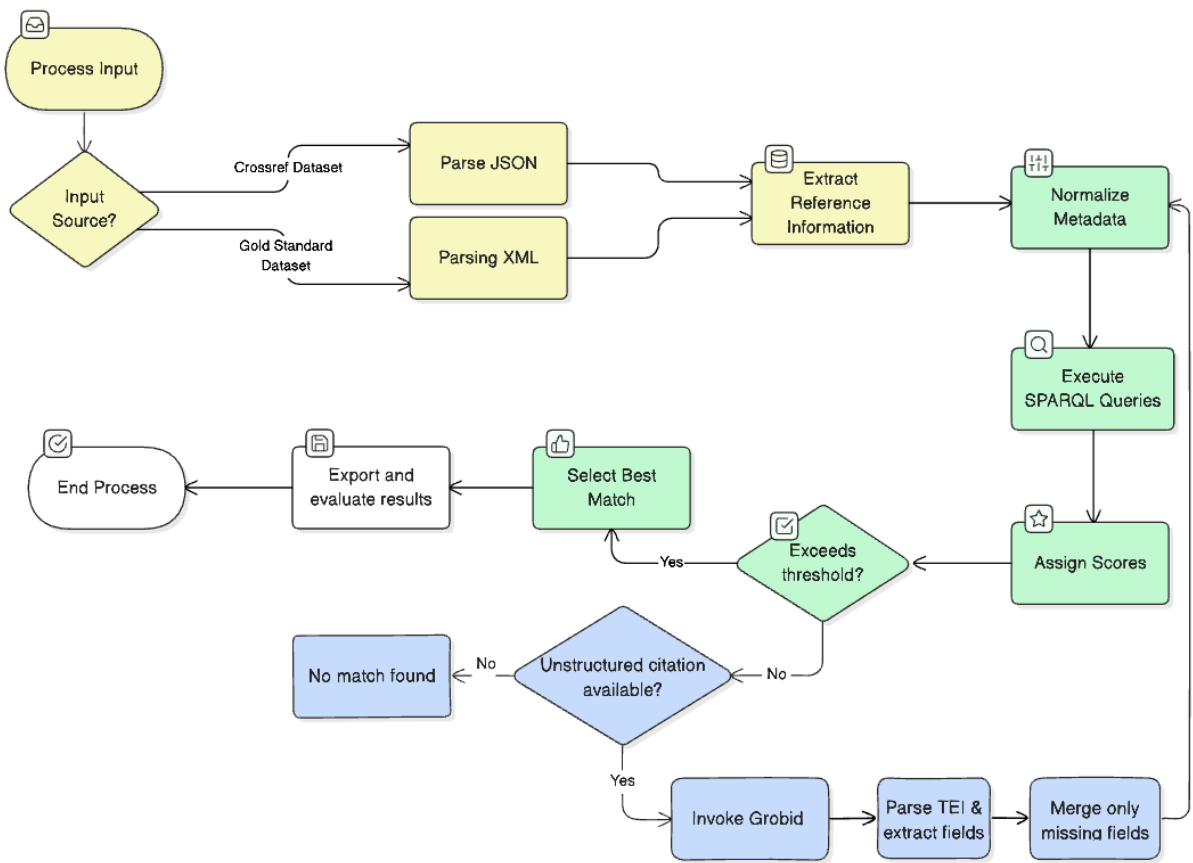

**Figure 1.** A workflow description of the four phases of the methodology used to analyse/identify incomplete bibliographic metadata. The four phases are: (yellow) data collection and preparation; (green) bibliographic metadata matching; (blue) enrichment fallback step for unstructured references; (white) results exportation and evaluation.

## 3.1 Data collection and preparation

This phase is dedicated to collecting and preparing all data required for the subsequent stages of the workflow. Specifically, it involves reading the input sources, the Crossref dataset and the Gold Standard, parsing their contents and extracting the reference data and metadata necessary for the analysis.

The Gold Standard is a manually annotated dataset that provides complete and accurate representations of bibliographic metadata. According to Pagnotta (2024), this dataset comprises 112 documents in XML-TEI format. The documents were selected to ensure coverage of 27 distinct academic disciplines and include Open Access papers, as well as a small subset of preprints. Each reference in the dataset is annotated with key bibliographic metadata, including author names, titles, journal or monograph information, publication year, volume and issue numbers, page ranges, publisher details, and DOIs when available. The annotation process followed the TEI standard, ensuring consistency with widely used bibliographic processing tools.

The Crossref data is retrieved via its open, free, and publicly accessible REST API, which returns results in JSON format. The service only supports Crossref DOIs (requests for non-Crossref DOIs return a 404 HTTP response). The Crossref API was queried by specifying each DOI from the articles included in the Gold Standard, using the following API command:

```
https://api.crossref.org/works/{doi}
```

The results include various types of metadata (institution, publication/registration dates, publisher, issue, ISSN/ISBN, licenses, funders, identifiers, authors, etc.). The attribute `reference` in the result returned by calling the Crossref API stores the reference list of the queried resource. Each reference (item of the list) is defined by a set of metadata attributes such as issue, DOI, volume, author, year, and journal title.

The Crossref dataset was constructed by retrieving metadata for the articles included in the Gold Standard. This was accomplished through manual DOI searches using the Crossref REST API. A key aspect of the Crossref dataset is its references section, where cited works are listed with varying degrees of metadata completeness. In some cases, references include structured metadata fields such as DOI, title, author names, volume, and pages, making them easily comparable to the Gold Standard. However, in many instances, references are stored as unstructured bibliographic reference strings, requiring additional processing to extract meaningful metadata. Furthermore, even structured references often lack crucial identifiers, such as DOIs, which significantly impairs the ability to establish reliable matches across databases.

The snippet of code in Listing 1 represents an XML-TEI reference in the Gold Standard, and its corresponding JSON representation in the Crossref database.

```xml
<biblStruct type="article" xml:id="b11">
    <analytic>
        <author>
            <persName>
                <surname>Forkman</surname>
                <forename>B</forename>
            </persName>
        </author>
        <author>
            <persName>
                <surname>Keeling</surname>
                <forename type="first">L</forename>
                <forename type="middle">J</forename>
            </persName>
        </author>
        <title level="a">
            Assessment of animal welfare measures for dairy cattle,
            beef bulls and veal calves
        </title>
    </analytic>
    <monogr>
        <title level="m">Welfare Quality® Reports</title>
        <imprint>
            <date when="2009">2009</date>
            <biblScope unit="volume">11</biblScope>
            <biblScope unit="page" from="297" to="" />
        </imprint>
    </monogr>
</biblStruct>
```

```json
{
  "first-page": "297",
  "article-title": "Assessment of animal welfare measures for dairy
                    cattle, beef bulls and veal calves",
  "volume": "11",
  "author": "Forkman",
  "year": "2009",
  "journal-title": "Welfare Quality® Reports"
}
```

**Listing 1.** A reference in the Gold Standard (top, in blue, XML format), and its corresponding JSON representation in the Crossref data (bottom, in black).

However, reference metadata in Crossref are often incomplete. As metadata are provided by different publishers with varying degrees of accuracy, the quality and completeness of the data can differ significantly. It is also likely to find unstructured metadata inserted in the dedicated field "unstructured", a possible example is shown in Listing 2.

```
"reference": [
    {
        "key": "ref1",
        "unstructured": "Badiene, A. (2013). Croissance Sans urbanisation
                         durable, pas de développement durable [Growth
                         without sustainable urbanization, no sustainable
                         development]. Jeune Afrique, 4(1), 41-47."
    }
    ...
]
```

**Listing 2.** Example of a reference entry in Crossref with the `unstructured` field specified.

The string in the `unstructured` field typically includes the title, authors, and year of publication, and may also contain additional details such as page numbers, volume, ISSN/ISBN, or DOIs. In the following section, we describe the methodology adopted to process these unstructured references, which relies on a distinct reference-matching workflow.

## 3.2 Bibliographic metadata matching

The matching process follows a cascading strategy, as detailed by Visser et al. (2021). The heuristic approach selects the best option at each step to obtain a global suboptimal solution, significantly reducing the number of options to explore (Hjeij, 2023). While this approach does not guarantee absolute optimality, it represents the fastest way to obtain accurate results with minimal computational effort.

During each phase, matches receive scores; ultimately, with an early stop mechanism, the first match exceeding the threshold is returned as an output that can be considered, with reasonable certainty, the correct match. This optimisation significantly reduces external API calls but may miss higher-scoring matches from less restrictive queries. If the threshold is not exceeded, a fallback strategy via GROBID (version 0.8.2) is activated to parse the unstructured reference and extract missing metadata, which is later merged and used to perform a new query. GROBID (GeneRation Of BIbliographic Data) is an open-source machine learning library that extracts, parses, and structures information from scholarly documents (mainly PDFs) into structured formats like TEI XML; it's widely used for automating citation and metadata extraction (Lopez, 2009; Grobid, 2008-2025).

The cascading process relies on SPARQL queries executed against OpenCitations Meta (https://w3id.org/oc/meta/sparql). When DOI information is available and the parameter *use_doi* of our matching tool is set to *True*, the system attempts DOI-based queries. The query cascade dynamically adapts based on available metadata, meaning that when a DOI is present in the reference and the *use_doi* parameter is enabled, the system follows the order presented in Table 1.

**Table 1.** Query cascading is used for matching bibliographic resources. Each rule executes SPARQL queries against OpenCitations Meta (https://w3id.org/oc/meta/sparql) and adapts to the available metadata. When *use_doi=True*, the cascade process prioritises DOI-based queries (Q1 and Q2).

| Title | Description |
|---|---|
| Q1-YEAR&DOI* | Matching resources with the same publication year and DOI. This criterion is the most reliable for accurate matching when both year and DOI are available, as it combines temporal and persistent identifier information.<br><br>* Is processed if the paprameter *use_doi=True* |
| Q2-DOI&TITLE* | Matching resources by DOI with the same title. It relies on the DOI and on the title of the article as a unique identifier, which is highly reliable despite the absence of temporal validation.<br><br>* Is processed if the paprameter *use_doi=True* |
| Q3-AUTH&TITLE | Matching resources with the same first author and title. Author-title matching provides a good balance between reliability and coverage. While author names and titles can have variations in their representation, they remain highly distinctive as identifiers when combined, especially after normalisation. |
| Q4-YEAR&AUTH&PAGE | Matching of resources with the same year, first author and starting page. This combination can capture matches missed by volume-based criteria, particularly in cases where volume information might be absent or inconsistently recorded. |
| Q5-YEAR&VOL&PAGE | Matching resources with the same publication year, volume number and starting page. This step relies on the assumption that the combination of these three metadata fields often uniquely identifies an article within a journal, particularly for journals with consistent volume and pagination practices. |
| Q6-YEAR&AUTH&VOL | Matching resources with the same publication year, first author and volume number. This different combination of fields already seen proved to be a good final fallback given the fairly unique nature of the three metadata fields combined together. |

While each query has specific mandatory fields for candidate retrieval, the comprehensive field set enables full scoring across all available metadata, improving match confidence even when only a subset is used for the initial search. The scoring system was designed to mirror the weighting scheme proposed by Visser et al. (2021) in proportion to the specific metadata fields used in our matching strategy. While Visser et al. (2021) assigned weights to a broader set of bibliographic fields, our implementation focuses on the core fields most consistently available across Crossref and OpenCitations data, with weights adjusted proportionally to maintain similar precision-recall trade-offs. The complete scoring breakdown is as follows:

- DOI exact match: 15 points

- Title matching:

    o Exact match (100% similarity): 14 points

    o Very high similarity (≥95%): 13 points

- o High similarity (≥90%): 13 points

- o Good similarity (≥85%): 12 points

- o Acceptable similarity (≥80%): 11 points

- o Moderate similarity (≥75%): 10 points

- Author exact match: 7 points

- Year matching:

  - o Exact year match: 1 point

  - o Adjacent year (±1 year): 0 points*

- Volume match: 3 points

- Page match: 8 points (combines validation of both starting and ending page numbers when available coming from the original scoring system in Visser, et al. (2021))

Adjacent year matches are tracked for analysis but do not contribute to the final score in the current configuration. The maximum possible score is 48 points (DOI + exact title + author + exact year + volume + page). The default acceptance threshold is set at 26 points, which represents approximately 54.5% of the maximum score, proportionally equivalent to the threshold used by Visser et al. (2021) in their matching strategy. This threshold can be customised, if needed, to accommodate different precision-recall trade-offs depending on the specific use case and data quality.

To avoid rejecting high-quality matches that fall marginally short of the acceptance threshold, the system implements an adaptive threshold adjustment mechanism. When the best candidate's score reaches at least 90% of the configured threshold, the effective threshold is dynamically lowered to match that candidate's score, allowing the match to be accepted. For example, with the default threshold of 26 points, a candidate scoring 23 points (90% of 26) or higher would trigger this adjustment, lowering the effective threshold to the candidate's score and permitting acceptance. This customisable mechanism helps capture legitimate matches that might otherwise be excluded due to minor metadata inconsistencies while still maintaining overall precision by requiring substantial agreement across multiple fields.

Initial SPARQL matching proceeds with original metadata only, but when a reference fails to exceed the threshold and contains an unstructured citation string, GROBID is invoked on-demand to extract additional metadata. The system then retries matching with the enriched reference. Experiments showed that this approach is more efficient in terms of recall and the results to be presented later (in Table 3) are coming from this configuration.

GROBID extracts year, author, title, volume, and page from unstructured strings. Year validation flags implausible values outside the interval *[1700, current_year+1]* as suspicious; if matching fails, the system retries with the year field cleared. Extracted values merge only into empty fields, preserving explicit metadata. When *use_doi=False*, the system skips DOI-based query strategies (Q1-YEAR&DOI and Q2-DOI&TITLE in Table 1) but still merges GROBID-extracted DOIs into the reference object. These DOIs contribute to match scoring via SPARQL OPTIONAL clauses in non-DOI queries, thereby improving match confidence without relying on DOI as the primary search criterion. This design allows the system to benefit from DOI information when available while still functioning effectively for references lacking DOI data. The *use_doi* flag thus controls query strategy rather than data availability: setting it to `False` means avoiding using DOI-based search strategies rather than ignoring DOIs completely. This distinction is important for maintaining high precision in scenarios where DOI coverage may be incomplete but other metadata fields are reliable. Comprehensive logging tracks extraction, merging, and retry outcomes for post-hoc analysis.

Single-file and directory processing generate matched references CSV (with reference ID, titles, score, DOI, Meta ID, query type) and one or more (depending on the input) unmatched references files documenting failures with metadata and best scores. The system implements asynchronous batch processing with configurable concurrency controls. When the input is a directory, files are processed in batches (default: 3 files per batch) with 10-second pauses between batches. Additionally, within each file, multiple references (up to 10 concurrently) are matched in parallel, enabling efficient SPARQL queries across both files and references. Incremental checkpointing (atomic writes every 10 files) enables automatic resumption after interruptions. Checkpoint files are also periodically compressed (every 10 batches) to prevent excessive growth during long-running operations.

The system leverages Python's asynchronous architecture to efficiently handle concurrent SPARQL queries via non-blocking HTTP requests with `aiohttp`. A semaphore limits concurrency to prevent endpoint overload, while a token bucket rate limiter regulates throughput at 2.5 requests per second with a burst capacity of 10, maintaining a balance between performance and endpoint stability. Both burst capacity and the requests per second are customisable. A more detailed workflow for the entire system is presented in Figure 2, with a special focus on the computation of the matching score.

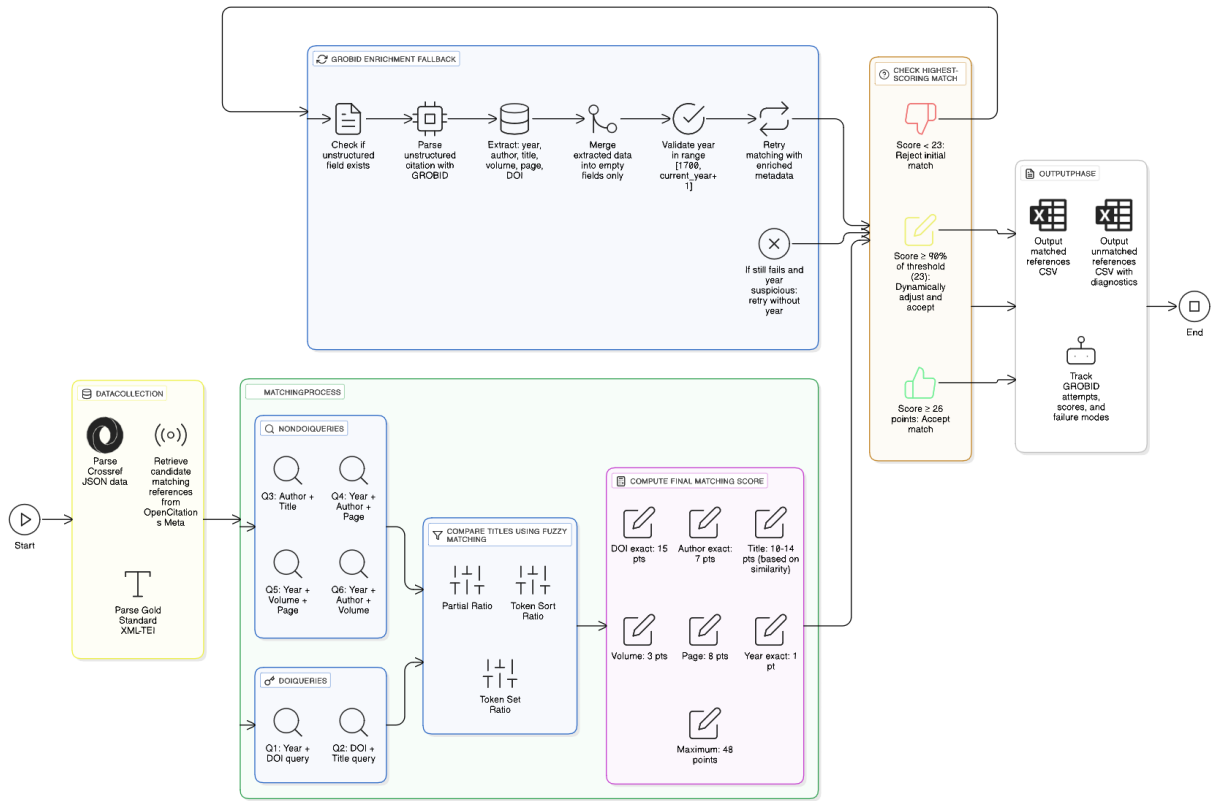

**Figure 2.** Diagram of the matching score computation process. The process starts by retrieving candidate references from OpenCitations Meta, followed by fuzzy title matching, year comparison with a ±1-year tolerance, and metadata normalisation. A final score is computed based on weighted criteria, with an early-stopping mechanism. If the threshold is not exceeded, a fallback strategy via GROBID is executed to parse the unstructured reference, extract missing metadata, and verify the results.

## 3.3 Handling errors and unstructured references

Beyond the core matching pipeline, the system provides comprehensive tracking of references that fail to produce successful matches. This functionality serves two purposes: first, it enables post-hoc analysis of matching performance by identifying which references lack sufficient metadata or fall outside OpenCitations Meta's coverage; second, it facilitates iterative refinement of the matching strategy by highlighting problematic cases.

The system tracks unmatched references in a dedicated CSV file containing comprehensive diagnostic information. Each unmatched record includes the reference identifier, all available metadata fields (year, volume, first page, first author surname, article title, volume title, journal title, DOI, and unstructured bibliographic reference string), along with detailed scoring information across multiple matching attempts. The system distinguishes between two failure types: (1) partial matches where candidates were found but scores fell below the threshold, which may indicate minor metadata inconsistencies, and (2) complete failures where no candidates were retrieved at all, suggesting either genuine absence from OpenCitations Meta or critically corrupted metadata fields.

For diagnostic purposes, the file records multiple score values when fallback mechanisms were activated: `score_original` captures the score achieved with original metadata, `score_after_grobid` records the score after GROBID enrichment (if attempted), and `score_without_year` shows the score achieved with the year field removed (if a suspicious year was found and removed). The `grobid_attempted` flag indicates whether GROBID fallback was triggered. This granular tracking enables analysis of fallback effectiveness and identification of cases where enrichment brought matches close to the threshold but did not exceed it, such as near-misses, which may indicate opportunities for threshold adjustment or metadata normalisation improvements. This distinction is critical for identifying whether failures stem from metadata inconsistencies or from a genuine absence in OpenCitations Meta, facilitating targeted refinement of the matching strategy.

Moreover, the system also implements a three-tier error handling strategy for SPARQL query failures:

- Rate Limiting (HTTP 429): Applies exponential backoff ($min(60, 2^{attempt} \times 5)$ seconds) with continuous token refill (2.5 tokens/second).
- Server Errors (HTTP 500-504): Triggers exponential backoff with jitter ($2^{attempt} + random(0,1)$ seconds). After 3 failed attempts, raises ServerError; in batch processing, 10 consecutive errors (configurable via `error_threshold`) trigger a 5-minute pause.
- Client Errors (HTTP 400-404): Retries up to 3 times with exponential backoff ($2^{attempt}$ seconds) before raising QueryExecutionError. Typically indicates SPARQL syntax errors or invalid identifiers.

## 4. Results

To evaluate the reference matching tool, a dedicated evaluation script was developed. This script implements three sequential operations: *check_doi*, *compare*, and *metrics*, which will be explained in detail as follows.

The first operation, called *check_doi*, reads JSON files containing Crossref metadata and verifies DOIs against OpenCitations Meta. The script takes a directory of JSON files as input and processes each file individually. For each JSON file, it extracts DOIs from the `message.reference` array, normalises them by converting to lowercase and unescaping forward slashes, then sends a SPARQL query to the OpenCitations Meta endpoint to check whether each DOI exists in their database.

The query returns bibliographic metadata if the DOI is found. The script classifies each DOI based on the query result and generates three output files per input file: a `*_doi_results.csv` file containing DOIs that returned results from OpenCitations Meta along with their metadata, a `*_unmatched_dois.csv` file containing DOIs that returned no results, and a `*_doi_statistics.csv` file with counts of total and successful queries.

The second operation, called *compare*, aims to compare two sets of DOIs on a per-file basis. It requires two input directories: one containing the `*_doi_results.csv` files generated by the *check_doi* operation, and another containing the `*_matches.csv` files produced by the reference-matching tool being evaluated.

The script matches files by their base filenames and compares the DOI sets in each pair. For each file pair, it identifies missed matches, which are DOIs present in the `*_doi_results.csv` file but absent in the `*_matches.csv` file, and earned matches, which are DOIs present in the `*_matches.csv` file but absent in the `*_doi_results.csv` file. These results are written to a single `comparison_results.csv` file that lists all missed and earned matches along with aggregate counts.

The final operation, called *metrics*, calculates classification metrics by treating OpenCitations Meta as the ground truth source. It requires the same two input directories as the compare operation: one containing the *check_doi* outputs and the other containing the reference-matching tool outputs.

The script processes files individually by loading three sets of DOIs for each file. The positive set (POS from now on) contains DOIs from `*_doi_results.csv`, representing DOIs that were found in OpenCitations Meta. The negative set (NEG from now on) contains DOIs from `*_unmatched_dois.csv`, representing DOIs that were not found in OpenCitations Meta. The predicted set (PRED from now on) contains DOIs from `*_matches.csv`, representing DOIs that were predicted by the reference matching tool being evaluated.

Using set operations, the script calculates four counts for each file. True Positives (TP) are calculated as the intersection of PRED and POS. False Positives (FP) are the intersection of PRED and NEG. False Negatives (FN) are POS minus PRED. True Negatives (TN) are NEG minus PRED. These counts are aggregated across all processed files.

From the aggregate counts, the script calculates standard classification metrics. Precision is computed as TP divided by the sum of TP and FP. Recall is TP divided by the sum of TP and FN. F1 Score is the harmonic mean of precision and recall. Accuracy is the sum of TP and TN divided by the sum of all four counts.

The script generates three outputs. An `overall_evaluation_metrics.csv` file contains the aggregate TP, FP, FN, and TN counts along with the calculated precision, recall, F1 score, and accuracy. The `metrics_debug_per_base.csv` file provides per-file counts for debugging.

A `filtered_matches/` directory contains CSV files with only the true positive matches for each file, combining metadata from both the `doi_results` and `matches` files.

The evaluation framework treats DOIs found in OpenCitations Meta as positive cases and DOIs not found in OpenCitations Meta as negative cases. The reference-matching tool's predictions are evaluated against this binary classification to assess performance.

In this evaluation, the Crossref JSON dataset consists of 100 files, while the Gold Standard XML dataset contains 112 files. This numerical discrepancy does not compromise the fairness of the evaluation, as the ground truth is established solely through the *check_doi* operation on the Crossref JSON files. The ground truth remains fixed and consistent across all evaluation phases, anchored to the DOIs extracted from the 100 JSON files and their verification status in OpenCitations Meta. The reference-matching tool processes the XML files and produces predictions, but only those corresponding to DOIs in the established ground truth are evaluated. Any matches produced by the tool from the 12 additional XML files that lack corresponding JSON files are excluded from the evaluation metrics, as they fall outside the defined ground truth scope. This approach ensures that all evaluated predictions are measured against the same consistent reference standard, maintaining the validity and comparability of the results.

Moreover, the pipeline distinguishes between processing failures and source files containing zero references. Files with 0 references in the source data (indicated by *reference-count=0* in Crossref JSON or absence of `biblStruct` elements in TEI XML) generate empty output files without triggering errors. These are tracked separately in aggregate statistics under the `empty_files` counter, distinct from files that failed due to processing errors.

This distinction is important for interpreting success rates: a file with 0 references is a successful processing outcome (correctly identifying no work to be done), not a failure requiring investigation. Aggregate reports show:

- *total_files_attempted*: All files processed;
- *files_processed*: Files containing ≥1 reference that completed successfully;
- *empty_files*: Source files with 0 references;
- *files_with_errors*: Files that failed due to processing errors.

It is worth noting that, as an output prior to evaluation, the pipeline produces a `.txt` file containing aggregate statistics to provide an effective overview of the match rate and the type of query that was most successful overall. In addition, an `.html` file is generated, allowing for a pleasant, immediate visualisation of the results of the analysis of the input data, as shown in Figure 3.

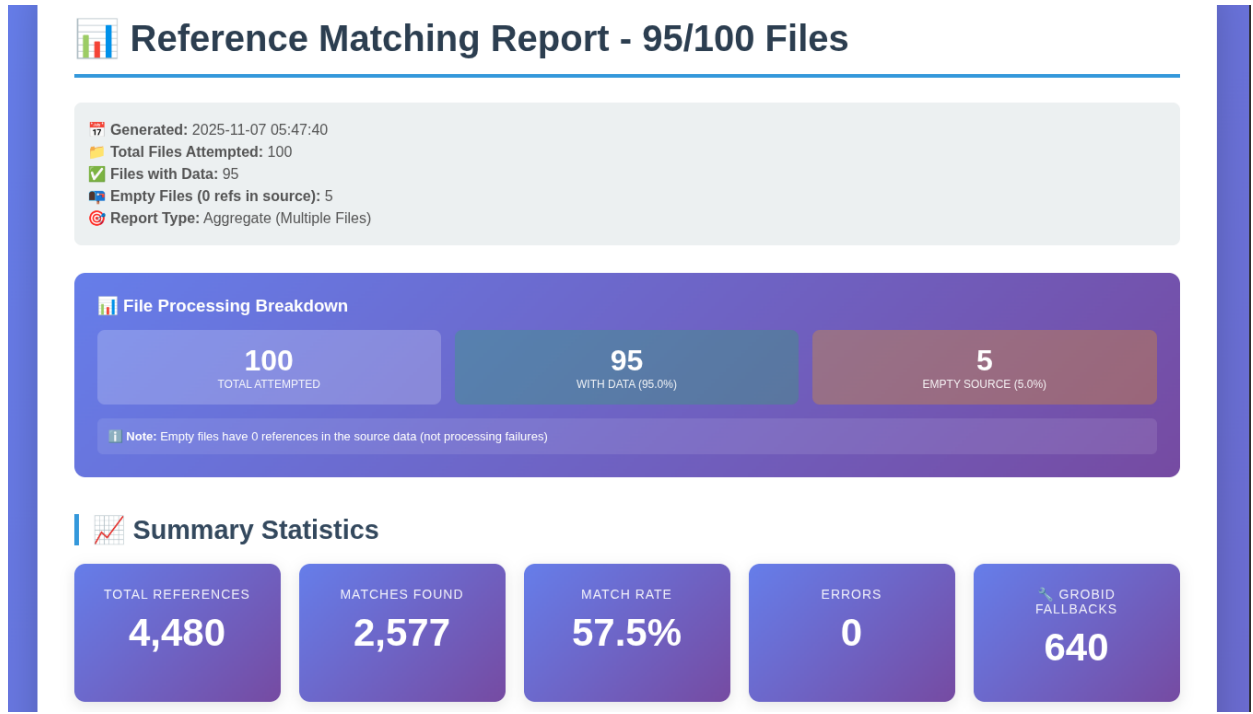

**Figure 3.** The HTML report immediately displays preliminary statistics belonging to the analysed data batch, in this case data from CrossRef obtained by the system with the parameter `use_doi=True`.

## 4.1 Gold Standard dataset

The evaluation conducted on the Gold Standard dataset yielded the results shown in Table 2.

**Table 2.** Gold Standard results. Results obtained by the developed matching tool on the Gold Standard dataset.

| METRIC | VALUE |
|---|---|
| **TRUE POSITIVES** | 2690 |
| **FALSE POSITIVES** | 0 |
| **FALSE NEGATIVES** | 865 |
| **TRUE NEGATIVES** | 6 |
| **PRECISION** | 100% |
| **RECALL** | 75.67% |
| **F1 SCORE** | 86.15% |

The high precision score indicates that the tool effectively minimises false positives, meaning that nearly all matches identified by the system were correct. However, the recall value suggests that some of the actual matches were not retrieved, highlighting the presence of false negatives. The

F1-score, which balances precision and recall, reflects an overall strong performance in this controlled setting.

## 4.2 Crossref dataset

The tool was evaluated on the Crossref dataset, the results are shown in Table 3.

**Table 3.** Crossref dataset results. Results obtained by the developed matching tool on the Crossref dataset.

| METRIC | VALUE (EXCLUDING DOI-BASED QUERIES) | VALUE (USING DOI-BASED QUERIES) |
|---|---|---|
| TRUE POSITIVES | 2073 | 2469 |
| FALSE POSITIVES | 0 | 0 |
| FALSE NEGATIVES | 1462 | 1086 |
| TRUE NEGATIVES | 6 | 6 |
| PRECISION | 100% | 100% |
| RECALL | 58.31% | 69.45% |
| F1 SCORE | 73.67% | 81.97% |

Like the Gold Standard dataset, the tool achieved 100% precision, indicating that no incorrect matches occurred. However, the recall value is lower, suggesting that many actual matches present in the dataset are not retrieved by the system. Consequently, the F1-score is also lower compared to the Gold Standard dataset, reflecting a trade-off between high precision and lower recall.

While the precision remains consistently high across both datasets, differences in recall suggest variations in the completeness of retrieved matches. It's also important to underline that 5 out of the 100 JSON files analysed for producing the results showed above were containing 0 references in the source Crossref data (reference-count: 0), not due to processing errors but because these papers legitimately lack bibliography metadata in Crossref's database (verified via direct Crossref API queries).

## 5. Discussions

This chapter will focus on analysing the key findings, highlighting the main achievements of the tool in different test scenarios. Additionally, the limitations of the approach will be critically examined, identifying areas for improvement that could enhance performance and reliability.

## 5.1 Key findings

The experimental evaluation revealed several important aspects of the tool's performance. One of the most notable findings is the exceptionally high precision achieved across both datasets – a

100% score for both the Gold Standard and Crossref datasets. This result demonstrates that the matching strategy effectively minimised false positives. The high precision is particularly relevant for bibliographic data integration, as reducing incorrect matches ensures the reliability of reference linking in scholarly databases.

However, the tool showed different recall levels across the two datasets, highlighting how metadata completeness and consistency affect performance. When tested on the Gold Standard dataset, the tool achieved a recall of 79.01%, resulting in an F1-score of 88.26%. These results indicate that, in a well-structured, well-annotated dataset, the tool successfully retrieved a large proportion of correct matches. Thus, the methodology works well when high-quality metadata are available.

In contrast, when evaluated on the Crossref dataset, recall dropped to 58.31%, yielding an F1-score of 73.67%. When `use_doi` was set to `True`, recall reached 69.45% and the F1 score was 81.97%. This difference can be attributed to the inherent limitations of metadata completeness in large-scale bibliographic databases such as Crossref. Given that Crossref metadata are often submitted by publishers with varying degrees of completeness, this result underscores the challenges of performing reliable reference matching in real-world scenarios.

The design choice to exclude DOIs from one of the two matching logic options allowed the tool to function even when DOI information was missing. However, this decision had significant implications for recall, as many references in Crossref are linked primarily via DOIs. While the tool identified matches across alternative metadata combinations, the evaluation was conducted only on references with DOIs verifiable through OpenCitations. This means that successful matches made by the tool for references without DOIs were not considered in the evaluation, potentially underestimating its recall.

Moreover, during the manual refinement phase of the tool, an important inconsistency in publication years between Crossref and OpenCitations Meta was frequently observed. Specifically, bibliographic resources that were otherwise identical often differed slightly in publication year between the two sources. This discrepancy is likely due to differences in how publication dates are recorded. Various stages in the publishing process can result in different values being assigned to the same resource across different databases. Recognising this issue led to implementing a ±1-year tolerance in the matching strategy, allowing a more flexible comparison while maintaining a high level of precision.

Overall, the findings indicate that while the tool demonstrates strong matching accuracy and precision, its ability to retrieve all possible correct matches is constrained by the availability and completeness of metadata. These insights highlight the importance of metadata quality in bibliographic databases and suggest potential areas for improvement in the tool's matching strategy.

## 6.2 Limitations

Despite its high precision, the tool has several limitations that should be considered when interpreting the results.

A major limitation lies in the evaluation methodology itself, particularly its reliance on DOIs to assess the correctness of matches. While the tool was designed to perform matching without requiring DOI information, the evaluation process relied exclusively on references with DOIs and verifiable in OpenCitations. This introduces several potential biases:

- *Exclusion of non-DOI references.* If the tool successfully matched a reference that did not contain a DOI in its metadata, that match was not considered in the evaluation. As a result, the recall metric may be lower than it would be in a real-world application where references without DOIs could still be correctly linked.

- *Dependence on OpenCitations' DOI coverage.* If a reference had a DOI but was missing from OpenCitations, it was excluded from the evaluation, even if the tool had successfully matched it. This could lead to an incomplete assessment of the tool's performance, particularly for references from less common or less frequently indexed sources.

- *Potential underestimation of recall.* Since the evaluation only included references with verifiable DOIs, it does not account for the tool's ability to match references based on alternative metadata. If a significant portion of references lacked DOI information but were still successfully matched, the actual recall could be higher than reported.

Moreover, while DOIs are the most common and reliable identifier in the datasets used for this study, Crossref also supports other persistent identifiers, such as ARKs (Archival Resource Keys), Handles, and ISSNs/ISBNs for journals and books. The decision to limit the matching process to DOIs was based on their prevalence in the selected dataset. However, a more comprehensive approach that incorporates additional identifiers could potentially improve the matching process.

Additionally, while the tool accounts for minor variations in titles, including different formatting of Greek letters and special characters, it does not fully address cases where the title or the authors' names are entirely written in a non-Latin script. In its current form, the matching system assumes that titles are either written in the Latin alphabet or transliterated into Latin characters. This poses a significant limitation when dealing with scientific literature published in languages using non-Latin scripts, such as Chinese, Arabic, Cyrillic, or Japanese. Future improvements could involve automatic transliteration mechanisms or multilingual text normalisation techniques to enhance compatibility.

## 6. Conclusions

This study provides a solid foundation for improving bibliographic reference matching in open citation databases. We have proposed a pipeline, based on existing heuristics and tools, to match bibliographic references with partial or no metadata against a collection of structured bibliographic information. In particular, we have experimented with bibliographic data from Crossref as defined in the field `reference`, which contains the list of bibliographic references of a given citing resource, to match them against OpenCitations Meta, the collection provided by OpenCitations, responsible for providing bibliographic metadata of all citing and cited entities referred to in the citations made available by the infrastructure. We have defined a precise methodology, implemented it in a software workflow, and run it against a manually defined Gold Standard and a subset of Crossref data to assess the quality of the results. While the heuristic-based tool developed has demonstrated strong matching precision and effective metadata integration, its recall limitations highlight the necessity of further enhancements to address metadata inconsistencies and leverage additional sources of citation data.

Beyond the scope of this research, the findings contribute to broader discussions on bibliographic data interoperability and citation network completeness. The challenges in matching references between Crossref and OpenCitations Meta reflect broader issues in scholarly metadata curation, underscoring the need for standardised, high-quality citation records to improve the accuracy of citation indexing and research assessment. Addressing these challenges will enhance the performance of bibliographic matching tools, support the growth of open citation infrastructures, and facilitate more transparent and accessible scholarly communication.

Anyway, the study presented in this article put the basis of already extending the existing coverage of OpenCitations data with missing citations that were not added in previous ingestion process since the Crossref data in the field `reference`, defining the cited resources of a given citing entity, did not specify any DOI as a structured field – the precondition in place of the current ingestion process run by OpenCitations. Considering that, as mentioned in the introduction, the number of references in the Crossref April 2025 dump are more than 698M and the heuristic-based tool was able to identify in OpenCitations Meta more than 58% of the cited entities described in the Crossref dataset used in the experiment, we can estimate we would be able to add to OpenCitations Index additional 404 million citation links (i.e. 58% pf 698M) that are not currently included in the infrastructure. In addition to expanding OpenCitations coverage as anticipated above, future research should focus on bridging the gap between heuristic and machine learning-based approaches, experimenting with customised versions of Grobid – e.g. (Soricetti, 2025) – and exploring hybrid strategies that balance precision, recall, and computational efficiency.

## Authors' contribution statement

Matteo Guenci: Software, Methodology, Validation, Writing – original draft, Writing – review & editing

Ivan Heibi: Data curation, Methodology, Supervision, Writing – original draft, Writing – review & editing

Chiara Parravicini: Software, Methodology, Writing – original draft, Writing – review & editing

Silvio Peroni: Conceptualization, Data curation, Funding acquisition, Methodology, Project administration, Resources, Supervision, Writing – original draft, Writing – review & editing

Marta Soricetti: Software, Writing – original draft, Writing – review & editing

## Acknowledgements

This work has been partially funded by the European Union's Horizon Europe framework programme under Grant Agreements No 101095129 (GraspOS), No 101188018 (GRAPHIA), and No 101187940 (LUMEN).

## Data availability statement

All the materials used in this study are openly available. Software, input data, and results are published on GitHub (https://github.com/opencitations/ref-matcher) and archived on Zenodo. The first upload was done on August 1, and the last update was on November 19 (Guenci et al., 2025).

## Conflict of interest

SP Director of OpenCitations. IH is the Chief Technology Officer of OpenCitations.